\documentclass[
superscriptaddress, twocolumn, 
reprint,
longbibliography,
superscriptaddress,
 amsmath,amssymb,
 aps,
prb,
]{revtex4-2}
\usepackage[T1]{fontenc}
\usepackage{graphicx}
\usepackage{dcolumn}
\usepackage{bm}
\usepackage{float}
\usepackage[colorlinks, linkcolor=red,citecolor=blue,urlcolor=blue]{hyperref}
\usepackage[all]{hypcap} 

\usepackage{xcolor}
\definecolor{darkblue}{rgb}{0,0,.65}
\definecolor{darkgreen}{rgb}{0.3,0.6,0.3}
\definecolor{darkorange}{rgb}{0.85,0.65,0.3}
\definecolor{cyan1}{rgb}{0.0, 0.6, 0.6}

\usepackage{soul}
\usepackage{ulem} 

\begin{document}

\preprint{APS/123-QED}

\title{Quadratic Hamiltonian approach to heat transport in fermionic systems}

\author{Ilari K. Mäkinen}
\affiliation{%
 Pico group, QTF Centre of Excellence, Department of Applied Physics,
Aalto University, P.O. Box 15100, FI-00076 Aalto, Finland
}%
\author{Ivan M. Khaymovich}%
\affiliation{
Nordita, Stockholm University and KTH Royal Institute of Technology Hannes Alfvens väg 12, SE-106 91 Stockholm, Sweden
}%
\author{Jukka P. Pekola}
\affiliation{%
 Pico group, QTF Centre of Excellence, Department of Applied Physics,
Aalto University, P.O. Box 15100, FI-00076 Aalto, Finland
}%

\date{\today}

\begin{abstract}
We investigate the problem of quantum heat transport, based on the quadratic fermionic systems with help of the Peschel trick of single-particle correlation functions. The efficient numerical method is applied to the particular case of a single mode heat valve and the results are compared to analytical formulae. Comparing several configurations and parameters we perform the systematic analysis of the method to most efficiently and accurately describe the simple quantum heat valve system.  
\end{abstract}

\maketitle


\section{Introduction}
\label{sec:introduction}
\begin{figure}
\centering
\includegraphics[width=\linewidth]{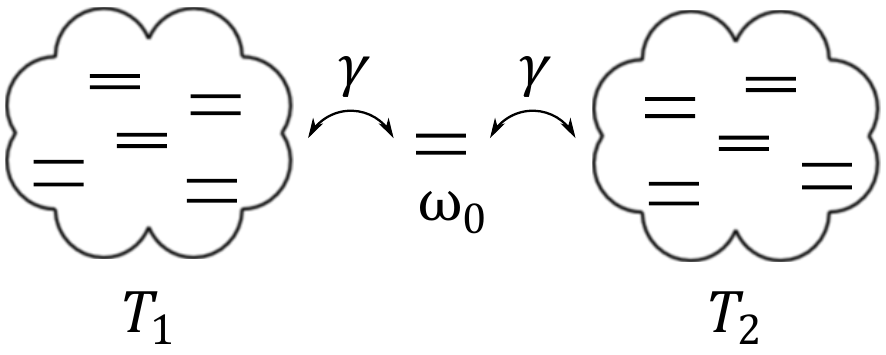}
\caption{The setup of the fermionic single mode heat valve. Two fermionic baths at respective temperatures $T_1$ and $T_2$ are connected through a single fermionic level of energy $\hbar\omega_0$. The parameter $\gamma$ is the characteristic coupling strength between the central level and the baths.}
\label{fig::fig0}
\end{figure}

The question of quantum heat transport and quantum thermalization has quite a long history. The main difference between classical chaotic systems and their quantum counterparts is that the latter obey significantly non-linear equations of motion, while the former satisfy fundamentally linear Schr\"odinger or von Neumann equations for the wave function or the density matrix, respectively.
As a result, classical chaos should not have a direct counterpart in the quantum world.

Already in 1949 A. Ya. Khinchin suggested the resolution of this problem. He invented the way to have thermalization in integrable (though classical) systems, obeying strictly linear equations of motion~\cite{khinchin1949mathematical}.
Later, in 1990s this idea led to the paradigm of eigenstate thermalization hypothesis (ETH) in quantum systems~\cite{Deutsch1991,Srednicki1994,Srednicki1996}.
The main idea behind this kind of quantum thermalization was that any local observable affects only a small part of the isolated (quantum) system, while the rest of it can work as a bath for this small measured part.

The crucial ingredient for the ETH was the scrambling over the nonlocal degrees of freedom which prevents one from detecting the deviations from thermalization by any local probe. This scrambling is based on the so-called volume law for the entanglement entropy, shown to be valid for the randomly distributed many-body quantum states~\cite{Page1993}, and crucially needs the presence of interactions in the quantum system.

However, recently several works have been done on the investigations of the generalization of ETH for local observables in single-particle~\cite{Vidmar-Rigol2021_PRB-2_1e_ETH} and many-body sectors~\cite{Vidmar-Rigol2023_PRL_genETH,Lydzba2024normal_weak_ETH}, as well as volume-law entanglement for translation-invariant~\cite{Vidmar-Rigol2017_PRL-1_TI_Fermi,Vidmar-Rigol2017_PRL-2_deviations} and local systems~\cite{Vidmar-Rigol2019_PRB_XY,Vidmar-Rigol2020_PRL_SYK2,Vidmar-Rigol2021_PRB-1_3d_ALT} in non-interacting quantum systems with random quadratic Hamiltonians.

There, based on the Peschel trick with the single-particle correlation function~\cite{Zanardi2002,Peschel2003,Calabrese2011,Peschel_2009}, the authors have calculated the upper bounds for the entanglement entropy for the random Slater-determinant states and demonstrated that even in the case of a quadratic Hamiltonian, it grows linearly with the volume of the system~\cite{Vidmar-Rigol2017_PRL-1_TI_Fermi}, considered the deviations from this maximum for the eigenstates of random quadratic Hamiltonians~\cite{Vidmar-Rigol2017_PRL-2_deviations,Vidmar-Rigol2020_PRL_SYK2}, XY-chains~\cite{Vidmar-Rigol2019_PRB_XY}, and 3d Anderson model~\cite{Vidmar-Rigol2021_PRB-1_3d_ALT}. 
Eventually they have shown that even if the eigenstates of quadratic chaotic Hamiltonians thermalize in a single-particle sector, their Slater determinant counterparts only equilibrate to the generalized Gibbs ensemble in the many-body one~\cite{Vidmar-Rigol2023_PRL_genETH}.

Recently, it has also been shown that the transport properties of the quadratic random quantum circuits show diffusive behavior~\cite{Dias2021} even without any conservation of the particle number.

All these recent developments on quantum thermalization in many-body non-interacting systems open an avenue for utilization of such methods for the quantum transport applications.

In this paper, we investigate the simplest, though experimentally feasible and relevant quantum-transport setup, see Fig.~\ref{fig::fig0}: namely, we consider two fermionic baths, each made of a large number of two-level systems (TLS) with distributed frequencies according to chosen densities of states $\nu_{\alpha}(\omega)$ for each bath $\alpha$, initialized with the Fermi-Dirac distributions at two different temperatures, and coupled to each other via a single two-level system, which approximates a qubit~\cite{Ronzani2018}, with fermionic statistics.
In such a setup, we test the above Peschel approach of the single-particle correlation function and the dependence of the quantum heat transport on various parameters:
size of the fermionic baths, particle conservation or its absence, weak, intermediate, or strong coupling of the qubit-like state to the reservoirs, and the internal couplings between the TLS in them.
The systematic analysis of all the above configurations and their comparison with the analytical prediction beyond the weak-coupling regime provides us with a deeper understanding of the protocols and parameters needed for the usage of quadratic fermionic Hamiltonians for the description of the experiments on quantum heat transport.

The rest of the paper is organized as follows.
In Sec.~\ref{sec:Method} we describe the general methodology of the calculations, based on quadratic fermionic Hamiltonians.
Sec.~\ref{subsec:coupling} is devoted to the application of the above method to the model in focus and Sec.~\ref{subsec:finte-size_effects} offers a systematic analysis of finite-size effects. Sec.~\ref{subsec:bath_coupling} discusses the role of internal bath couplings between TLS. Sec.~\ref{sec:conclusion} concludes our analysis and offers discussion on further applications of the method.

\section{Peschel trick of single-particle correlation functions}\label{sec:Method}
A general operator $\mathcal{O}$ quadratic in the fermionic creation and annihilation operators can always be written in the particle-hole symmetrized form 
\begin{equation}
    \mathcal{O} = \frac{1}{2}\vec{A}^{\dagger}O\vec{A} + \mathrm{const.},
\label{eq::nambu_form}
\end{equation}
where the vector $\vec{A} = (a_1,\dots, a_N, a_1^{\dagger}, \dots, a_N^{\dagger})^{T}$ and the $2N\times2N$ matrix $O$ is the Nambu representation of the operator. In particular, we focus on quadratic fermionic Hamiltonians $\mathcal{H}$, which can always be diagonalized by defining fermionic quasiparticle modes via a unitary transformation $U$ such that~\cite{Colpa1978}
\begin{equation}
\label{eq::boguliubov}
    \mathcal{H} = \frac{1}{2}\vec{A}^{\dagger}H \vec{A} = \frac{1}{2}\vec{B}^{\dagger}D\vec{B},
\end{equation}
where $D = U^\dagger HU$ is a diagonal matrix and the vector $\vec{B}=U^\dagger \vec{A}$ collects the annihilation and creation operators of the quasiparticle modes. In particular, given the diagonalized form, the Heisenberg equation of motion for the mode vector $\vec{B}$ simply reads as
\begin{equation}
    \hbar\partial_t \vec{B} = -i D\vec{B}
\end{equation}
and by transforming back to the original mode structure, we see that similarly
\begin{equation}
    \hbar\partial_t\vec{A}=-i(UDU^\dagger)\vec{A} = -iH\vec{A}.
\end{equation}
The time-evolution of the single-particle correlation matrix $\chi(t)=\langle \vec{A}(t) \vec{A}^{\dagger}(t)\rangle$ is thus given by
\begin{equation}
\label{eq::correlation-evolution}
\begin{split}
    \chi(t)
    &=e^{-iHt/\hbar}\langle \vec{A}(0) \vec{A}^\dagger(0)\rangle e^{iHt/\hbar} \\&= e^{-iHt/\hbar}\chi(0)e^{iHt/\hbar}.   
\end{split}
\end{equation}
This formula allows us to compute the time evolution of the single-particle correlation matrix given the representation matrix $H$ of the quadratic fermionic Hamiltonian and the initial correlation matrix $\chi(0)$. In the case of a bosonic quadratic Hamiltonian, the generalized Bogoliubov transformation used to obtain the diagonalization in Eq.~\eqref{eq::boguliubov} does not necessarily succeed because the conditions for a representation in terms of non-interacting bosonic quasiparticles require more than just the Hermiticity of the representation matrix $H$~\cite{Colpa1978}. Therefore, here and further, we focus on the fermionic TLS modes to model the baths and the qubit.

Given the time-evolution of the single-particle correlation matrix, the time-evolution of the expectation value $\langle \mathcal{O}\rangle$ of a quadratic fermionic operator is then computed by taking the trace~\cite{Dias2021}
\begin{equation}
\begin{split}
     \langle \mathcal{O}\rangle (t) &= \mathrm{tr}(\mathcal{\mathcal{O}}\rho)
    =\frac{1}{2}\sum_{i,j}O_{ij}\mathrm{tr}(A_i^{\dagger}A_j\rho)+\mathrm{const.}  
    \\&=
    -\frac{1}{2}\sum_{i,j}O_{ij}\langle  A_jA_i^\dagger\rangle+\mathrm{const.}  
    \\&=
    -\frac{1}{2}\mathrm{tr}(O\chi(t))+\mathrm{const.},
\end{split}
\end{equation}
where the fermionic anticommutation relation was used to exchange the i-th and the j-th elements $A_i^\dagger$ ($A_j$) of the vector $\vec{A}^{\dagger}(t)$ ($\vec{A}(t)$), given by $a_i^\dagger$ ($a_j$) for $i,j\leq N$ and $a_{i-N}$ ($a_{j-N}^\dagger$) otherwise.
Moreover, from the last two equations, it follows that the time-derivative of the expectation value $\langle\mathcal{O}\rangle(t)$ can be calculated with the equation
\begin{equation}
    \frac{d}{dt}\langle\mathcal{O}\rangle(t) = -\frac{1}{2i\hbar}\mathrm{tr}(\chi(t)[O,H]).
\label{eq::current}
\end{equation}
In the particular case, where $\mathcal{O}$ is the Hamiltonian for a part of the total system, this expression captures the heat current flowing into that part. 

The method has previously been applied in the study of random fermionic circuits~\cite{Dias2021}, where it served to translate computations with many-body observables into tractable single-particle problems. Regarding the applicability of the method to the modeling of heat transport, its main limitation is the necessity of a quadratic Hamiltonian, and therefore it cannot be used for capturing non-linear transport phenomena such as rectification~\cite{Segal2005, Senior2020}. The applicability of the method to non-linear systems such as qubits is limited to their linear regime. Nevertheless, the physics captured by the linear analysis includes interesting information regarding, for example, the tunability of the heat current by an external parameter~\cite{Teemu2008, Satrya2023}
and the low-temperature behavior of heat conductance \cite{Agarwalla2017}. 
\section{Application to a single mode heat 
valve}\label{subsec:coupling}
\begin{figure}
\centering
\includegraphics[width=\linewidth]{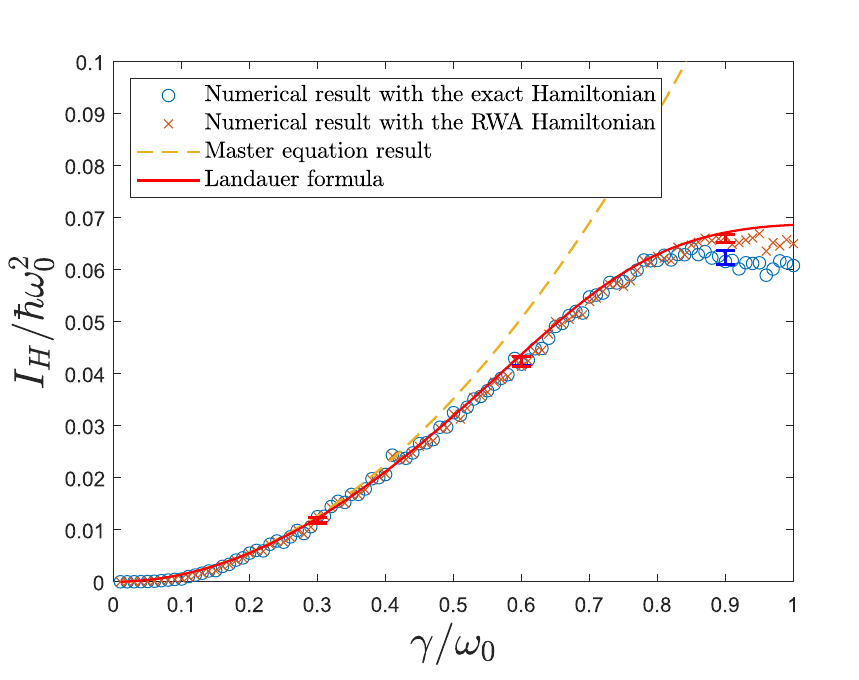}
\caption{The estimated steady-state heat current through the single level fermionic heat valve as a function of the relative coupling strength $\gamma/\omega_0$ computed by averaging the simulated heat current $I_H(t)$ (Eq.~(\ref{eq::heat_current_to_cold})) into the cold bath on the time-interval $\omega_0t \in [20,50]$. The simulation was performed for both the exact Hamiltonian of~Eq.~(\ref{eq::full_hamilton}) and the RWA Hamiltonian of Eq.~(\ref{eq::rwa_hamilton}) with the each of two baths containing $N=1200$ fermionic oscillators. The error bars correspond to the standard deviations resulting from running the simulation procedure multiple times. The randomness is due to the generated bath frequencies and the couplings.  The initial temperatures of the hot and the cold bath are $T_1 = \hbar\omega_0/k_{\mathrm{B}}$ and $T_2 = 0$ respectively. The parabolic dashed line shows the weak coupling prediction and the solid line is the prediction of a Landauer type formula~\cite{Topp_2015} (see Appendix~\ref{app::steady-state}).  }
\label{fig::fig2}
\end{figure}
To apply the single-particle correlation function method described in the previous section to the single mode heat valve introduced in Sec.~\ref{sec:introduction} and illustrated in Fig.~\ref{fig::fig0}, we need to specify the total Hamiltonian $\mathcal{H}$, the Hamiltonians $\mathcal{H}_\alpha$ of the baths, and the initial single-particle correlation matrix $\chi(0)$. The total Hamiltonian reads
\begin{equation}
\begin{split}
    \mathcal{H}/\hbar = &\sum_{\alpha,k}\omega_{\alpha k}a^\dagger_{\alpha k}a_{\alpha k} + \omega_0d^\dagger d+\\&\sum_{\alpha,k} g_{\alpha k}(a_{\alpha k}^{\dagger}d+d^\dagger a_{\alpha k} + a_{\alpha k}^\dagger d^\dagger+da_{\alpha k}),     
\end{split}  
\label{eq::full_hamilton}
\end{equation}
where $\{\hbar\omega_{\alpha k}\}$ are the energies labeled by $k = 1,2,\dots, N$ in the baths $\alpha \in \{1,2\}$, $\hbar\omega_0$ is the energy of the qubit-like TLS between the two baths and $\{g_{\alpha k}\}$ are the couplings between the qubit-like TLS and the fermionic bath oscillators. The Hamiltonian of the bath $\alpha \in \{1,2\}$ 
\begin{equation}
\label{eq::bath_hamiltonian}
    \mathcal{H}_{\alpha}/\hbar
    =
    \sum_k\omega_{\alpha k}a_{\alpha k}^\dagger a_{\alpha k}.
\end{equation}
In addition, we also study the corresponding Hamiltonian in the rotating-wave approximation given by
\begin{equation}
\begin{split}
    \mathcal{H}_{\mathrm{RWA}}/\hbar = &\sum_{\alpha,k}\omega_{\alpha k}a^\dagger_{\alpha k}a_{\alpha k} + \omega_0d^\dagger d+\\&\sum_{\alpha,k} g_{\alpha k}(a_{\alpha k}^{\dagger}d+d^\dagger a_{\alpha k}),     
\end{split}  
\label{eq::rwa_hamilton}
\end{equation}
where we have simply dropped the particle non-conserving terms of the exact Hamiltonian.

Given the total Hamiltonian $\mathcal{H}$ or $\mathcal{H}_\mathrm{RWA}$, we still have the freedom to choose the distribution $\nu_{\alpha}(\omega)$ of the bath energies $\{\hbar\omega_{\alpha k}\}$ and the couplings $\{g_{\alpha k}\}$. To illustrate the method, we use a uniform distribution on the interval $[0,2\omega_0]$ from which we generate $N=1200$ energies for each of the two baths, and the couplings to the single level are generated from a likewise uniform distribution on the interval $[-\gamma/\sqrt{N},\gamma/\sqrt{N}]$, where $\gamma$ is a parameter that determines the characteristic coupling strength between the qubit-like TLS and the baths.

The initial conditions are chosen such that the two baths are disentangled from the qubit-like TLS with their mode populations according to the Fermi-Dirac distribution at the respective temperatures $T_1$ and $T_2$, and the single mode has zero initial population.

Hence, the initial single-particle correlation matrix is simply diagonal, collecting all the initial populations of the bath modes and the qubit-like TLS. To give an example of the simulated heat current, the temperature of the hot bath is chosen to be $T_1 = \hbar\omega_0/k_B$ in order to achieve reasonably large populations near the central TLS energy and the temperature of the cold bath is chosen to be $T_2 = 0$ for convenience.  Finally, using the respective Nambu representation matrices $H$ and $H_2$ of the exact total Hamiltonian and the cold bath, we can apply Eq.~(\ref{eq::current}) to write the heat current $I_H(t)$ into the cold bath as
\begin{equation}
\label{eq::heat_current_to_cold}
    I_H(t) = -\frac{1}{2i\hbar}\mathrm{tr}(\chi(t)[H_2,H]),
\end{equation}
which is then evaluated numerically.
The practical results of applying this computation procedure are best illustrated by plotting the obtained numerical estimate of the steady-state heat current at long times after the transient processes are gone as a function of the relative coupling strength $\gamma/\omega_0$ as shown in Fig.~\ref{fig::fig2}, where the numerical result is also compared with the result of the master equation valid for weak coupling, and the Landauer formula valid for general coupling~\cite{Topp_2015} (see Appendix \ref{app::steady-state}). Additionally, these two formulas, used to compare the numerical results with, assume the large bath size limit, where the bath frequency distributions are well-approximated by continua. Thus, we naturally expect differences to these formulae due to finite-size effects.

Furthermore, we also show the corresponding result for the rotating-wave approximation (RWA), where the Hamiltonian is given by Eq.~(\ref{eq::rwa_hamilton}). The result differs only slightly from that of the exact Hamiltonian, as one would expect, since the rotating-wave approximation is increasingly valid at long time scales at which the steady state is reached. 

As the coupling is increased while keeping the bath sizes fixed, the two numerical steady-state estimates start to differ significantly from the Landauer formula as already seen from Fig.~\ref{fig::fig2}. The difference is caused by finite size effects. Firstly, a finite bath cannot sustain its thermal distribution, and thus the hot bath is exhausted and the cold bath is overpopulated near the frequency of the central TLS before the system has had time to approach the continuum steady state of the Landauer formula, which implies that the numerical estimates tend to the continuum limit from below as the bath size is increased. Secondly, for small bath sizes or strong enough couplings, the simulated heat current shows significant oscillations on the chosen averaging interval and, therefore, the heat current is no longer close to a steady state. These effects are further elaborated on and studied systematically in the next section.

\section{Finite size effects and the rotating-wave approximation}
\label{subsec:finte-size_effects}
\begin{figure}
\centering
\includegraphics[width=\linewidth]{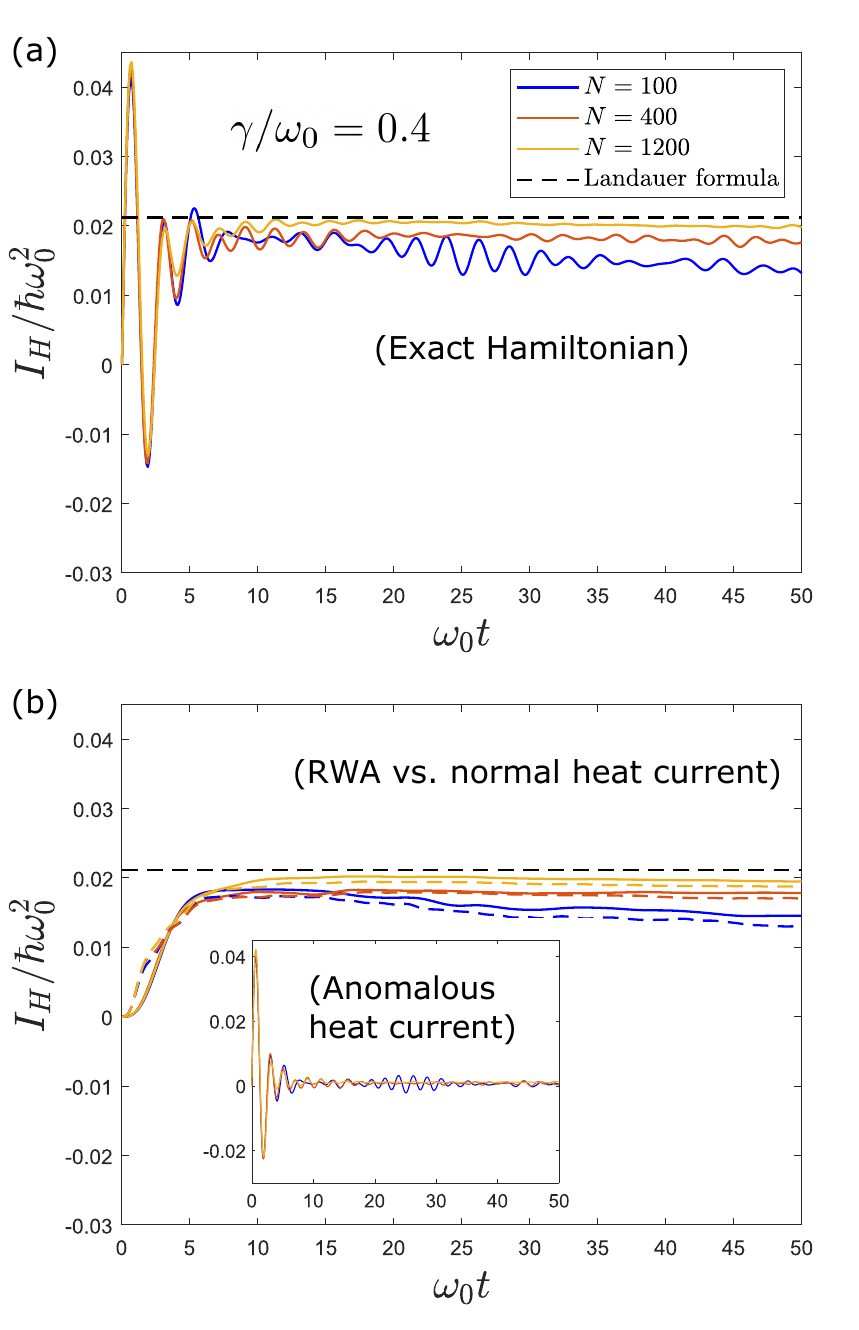}
\caption{(a) The total heat current into the cold bath (Eq.~(\ref{eq::heat_current_to_cold})) as a function of time computed with the exact Hamiltonian with different bath sizes $N$ and a common coupling strength $\gamma/\omega_0 = 0.4$. The dashed line is the steady-state prediction according to a Landauer type formula (see Appendix~\ref{app::steady-state}). (b) The corresponding result computed with the RWA Hamiltonian is shown by the solid lines. Moreover, the dashed lines show the normal part of the exact heat current (Eq.~(\ref{eq::norm_and_anom_currents})) from (a) and the inset shows the anomalous part (Eq.~(\ref{eq::norm_and_anom_currents})). This demonstrates that the oscillations of the exact heat current are due to the anomalous part, whereas the steady state is mostly determined by the normal part.}
\label{fig::fig3}
\end{figure}

The steady-state result discussed in the previous section appears only in the limit of large bath sizes. Moreover, it obviously takes a finite time for the heat current to become sufficiently stable, during which the rotating-wave approximation will not agree with the exact time-evolution. An illustration of these effects is shown in Fig.~\ref{fig::fig3}, where we plot both the exact total heat current and the rotating-wave approximation as a function of time for multiple bath sizes with a fixed relative coupling strength $\gamma/\omega_0=0.4$. The horizontal dashed line is the steady-state prediction of the Landauer formula. 
With increasing system size, the exact heat current in Fig.~\ref{fig::fig3}(a) shows less and less deviations from the Landauer formula, confirming that this deviation is due to the finite-size effects.

\begin{figure}
\centering
\includegraphics[width=\linewidth]{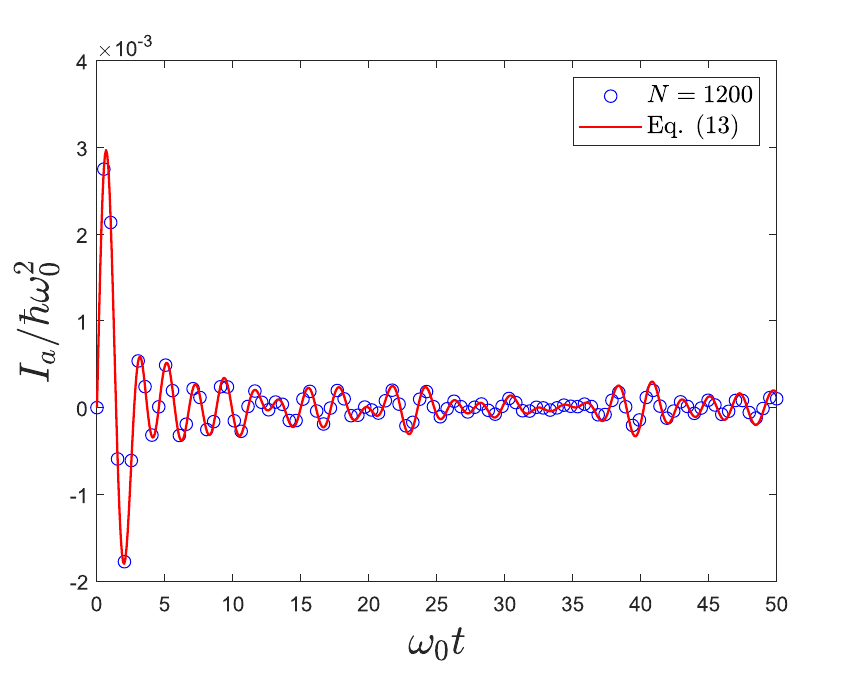}
\caption{The computed anomalous heat current (Eq.~(\ref{eq::norm_and_anom_currents})) into to the cold bath as a function of time at the coupling strength $\gamma/\omega_0 = 0.1$ compared with the approximate formula given by Eq.~\eqref{eq::anomalous_current}.}
\label{fig::fig4}
\end{figure}
Furthermore, the result confirms that the rotating-wave approximation fails to capture the significant oscillations present in the exact heat current. These oscillations in the exact result are nevertheless damped with time, which eventually causes the rotating-wave approximation to be in good agreement with the exact result. In the regime of sufficiently weak coupling, this transient effect is conveniently explained by writing the total heat current $I_H$ as a sum of two contributions
\begin{equation}
\label{eq::norm_and_anom_currents}
\begin{split}
    I_H(t)&=I_n(t)+I_a(t)\\&=\underbrace{-i\hbar\sum_{k} g_{\alpha k}\omega_{\alpha k} (\langle a_{\alpha k}^\dagger d\rangle - \langle d^\dagger a_{\alpha k}\rangle)}_{I_n(t)}\\&~~~~\underbrace{-i\hbar\sum_{k} g_{\alpha k}\omega_{\alpha k} (\langle a_{\alpha k}^\dagger d^\dagger\rangle - \langle d a_{\alpha k}\rangle)}_{I_a(t)},
\end{split}
\end{equation}
where we call $I_n(t)$ and $I_a(t)$ as the normal and the anomalous current, respectively. Computing these contributions separately yields the results shown in Fig.~\ref{fig::fig3}(b), which demonstrates that the significant oscillations of the total heat current are due to the anomalous part (see the inset), whereas the normal part shows approximately similar behavior to the total heat current computed in the RWA. Using the weak coupling assumption $\gamma/\omega_0\ll1$, the analysis can be quantified further by studying the approximate expression (see Appendix~\ref{app:anomalous_current})
\begin{equation}
\label{eq::anomalous_current}
    I_a(t) \approx 2\hbar\langle g_{\alpha k}^2\rangle\sum_{k}\frac{\omega_{\alpha k}[1-f(\beta_\alpha\hbar\omega_{\alpha k})]}{\omega_0+\omega_{\alpha k}}\sin((\omega_0+\omega_{\alpha k})t).
\end{equation}
 The comparison with the numerically obtained exact result is shown in Fig.~\ref{fig::fig4}, which demonstrates the transient nature of the anomalous current term. Moreover, in the large bath size limit, when each bath frequency realization is well-approximated by the continuous distribution, we have
\begin{equation}
    I_a(t) \approx 2\hbar\langle g_{\alpha k}^2\rangle\int d\omega\nu_{\alpha}(\omega)\frac{\omega[1-f(\beta_\alpha\hbar\omega)]}{\omega_0+\omega}\sin((\omega_0+\omega)t),
\end{equation}
which clarifies the role of the bath density of states $\nu_\alpha(\omega)$ as a weighting factor in the integral over the energies.

\section{Internal bath couplings}
\label{subsec:bath_coupling}
The bath Hamiltonians defined by Eq.~(\ref{eq::bath_hamiltonian}) consist of non-interacting fermionic oscillators. However, the framework of quadratic Hamiltonians still allows us the freedom to add quadratic terms, that couple the oscillators with each other. The introduction of these internal bath couplings corresponds to a change in the bath eigenmodes and the spectrum. Physically, such couplings could correspond to internal relaxation mechanisms or to a tunable environment.  Assuming that the internal couplings are quadratic and involve only particle conserving terms for simplicity, the diagonalization of a bath means that we can find a unitary matrix $U^{(\alpha)}$ in terms of which the bath Hamiltonian $\mathcal{H}^{(\alpha)}$ is diagonalized as
\begin{equation}
    \mathcal{H}^{(\alpha)}
    =
    \vec{A}^{(\alpha)\dagger} H^{(\alpha)}\vec{A}^{(\alpha)}
    =
    \vec{B}^{(\alpha)\dagger}D^{(\alpha)}\vec{B}^{(\alpha)},
\end{equation}
 where $\vec{A}^{(\alpha)} = (a_{\alpha 1}, \dots,a_{\alpha N})^{T}$, $D^{(\alpha)}$ is a diagonal matrix and $\vec{B}^{(\alpha)} = U^{(\alpha)\dagger}\vec{A}^{(\alpha)}$. Note that this description is different from the one introduced in Sec. \ref{sec:Method}, which is made possible by involving only particle-conserving terms. Of course, one could still work with the previous description, but then the Nambu representation matrix of the Hamiltonian would be block diagonal with the two blocks differing only by a minus sign. Hence, it is sufficient to work with only $N\times N$ matrices, in contrast to $2N\times 2N$ matrices of the full description. 
 
 Following the introduced notation, the coupling Hamiltonian to the central level is modified according to
\begin{equation}
\begin{split}
     V/\hbar &= \sum_{\alpha,k}g_{\alpha k}(a_{\alpha k}^\dagger d+d^\dagger a_{\alpha k}) 
     \\&= \sum_{\alpha,j,k} g_{\alpha k}(U_{kj}^{(\alpha)*}b_{\alpha j}^{\dagger}d+U_{kj}^{(\alpha)}d^{\dagger}b_{\alpha j})
     \\& = \sum_{\alpha,j}\left(\sum_k U_{kj}^{(\alpha)*}g_{\alpha k}\right)b_{\alpha j}^{\dagger}d+\left(\sum_k U_{kj}^{(\alpha)}g_{\alpha k}\right)d^{\dagger}b_{\alpha j}.
\end{split}
\end{equation}
The above expressions state that the bath energies are shifted to new values given by the diagonal values of the matrix $D^{(\alpha)}$ and the couplings to the qubit-like TLS are transformed according to $g_{\alpha j}\rightarrow \sum_k U_{k j}^{(\alpha)}g_{\alpha k}$. However, only the former effect is important because the unitary transformation of the couplings conserves the expectation value $\langle |g_{\alpha k}|^2\rangle\equiv \langle (1/N)\sum_k|g_{\alpha k}|^2\rangle$, which, according to the Landauer formula (see Appendix~\ref{app::steady-state}), is the only contribution of the coupling distribution to the steady-state heat current. 
\begin{figure}
\centering
\includegraphics[width=\linewidth]{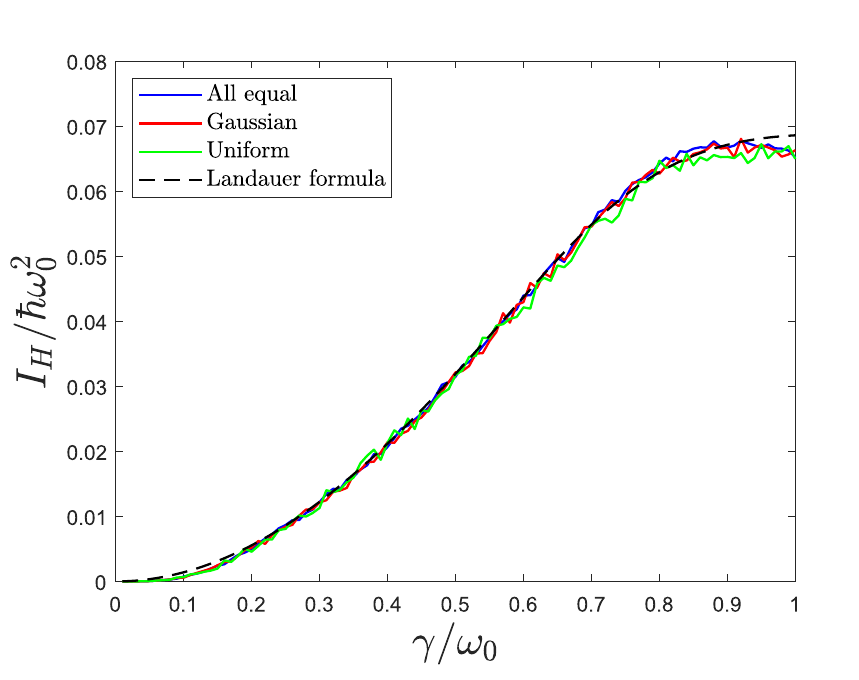}
\caption{The estimated RWA steady-state heat current through the single level fermionic heat valve as function of the relative coupling strength for three different coupling distributions. All the distributions share the same second moment. The number number of fermionic oscillators in each of the two baths are set to $N=2000$. The initial temperatures of the hot and the cold bath are $T_1 = \hbar\omega_0/k_{\mathrm{B}}$ and $T_2 = 0$ respectively.  }
\label{fig::fig5}
\end{figure}
The new single-particle bath energies, on the contrary, appear explicitly in the Landauer formula through the density of states. Therefore, it is sufficient to account for the internal couplings by simply replacing the old single-particle bath energies by the new ones due to the effect of internal couplings. This simplifies the simulation procedure by making the representation matrix of the total system mainly diagonal.

As an example to demonstrate the irrelevance of the precise coupling distribution, let us reconsider the steady-state simulation of Sec.~\ref{subsec:coupling} with different distributions of the couplings to the qubit-like TLS. Specifically, in Fig.~\ref{fig::fig5}, we show the estimated steady-state heat current as a function of the relative coupling strength for 
the original uniform distribution used in Sec.~\ref{subsec:coupling}, the zero mean Gaussian one, and deterministic all-to-all equal couplings. 
All three coupling distributions are set to have equal second moments. As expected, all three steady-state estimates are in good agreement with each other. 

\section{Conclusion}\label{sec:conclusion}
Our simulation results show that the single-particle correlation function method can be successfully applied to study heat transport in quadratic fermionic systems containing arbitrary coupling strengths and finite temperatures. In particular, the method can be used to study how the approach to steady state depends on various parameters such as the system size and the initial one-particle correlation matrix. We focused on an initial state without correlations between the central mode and the baths, but further simulations could be used to understand the effect of such correlations and the implications of the RWA in these setups.

Obviously, the structure connecting the heat baths is not limited to the simplest case of a single mode studied here. Possible interesting structures include, for example, longer chains of fermions or parallel transport structures. With modifications of the observable, the method can also be used to simulate any quadratic observable such as the matter current, and therefore one could for instance study the differences between the matter and heat currents in various parameter regimes.

\acknowledgments
IKM acknowledges the financial support of the Finnish Ministry of Education and Culture through the Quantum Doctoral Education Pilot Program (QDOC VN/3137/2024-OKM-4) and the Research Council of Finland through the Finnish Quantum Flagship project (Aalto 358877). IMK acknowledges support by the European Research Council under the European Union’s Seventh Framework Program Synergy ERC-2018-SyG HERO-810451. This work was financially supported by the Research Council of Finland Centre of Excellence program grant 336810 and grant 349601 (THEPOW).

\bibliography{apssamp}

\begin{thebibliography}{26}%
\makeatletter
\providecommand \@ifxundefined [1]{%
 \@ifx{#1\undefined}
}%
\providecommand \@ifnum [1]{%
 \ifnum #1\expandafter \@firstoftwo
 \else \expandafter \@secondoftwo
 \fi
}%
\providecommand \@ifx [1]{%
 \ifx #1\expandafter \@firstoftwo
 \else \expandafter \@secondoftwo
 \fi
}%
\providecommand \natexlab [1]{#1}%
\providecommand \enquote  [1]{``#1''}%
\providecommand \bibnamefont  [1]{#1}%
\providecommand \bibfnamefont [1]{#1}%
\providecommand \citenamefont [1]{#1}%
\providecommand \href@noop [0]{\@secondoftwo}%
\providecommand \href [0]{\begingroup \@sanitize@url \@href}%
\providecommand \@href[1]{\@@startlink{#1}\@@href}%
\providecommand \@@href[1]{\endgroup#1\@@endlink}%
\providecommand \@sanitize@url [0]{\catcode `\\12\catcode `\$12\catcode `\&12\catcode `\#12\catcode `\^12\catcode `\_12\catcode `\%12\relax}%
\providecommand \@@startlink[1]{}%
\providecommand \@@endlink[0]{}%
\providecommand \url  [0]{\begingroup\@sanitize@url \@url }%
\providecommand \@url [1]{\endgroup\@href {#1}{\urlprefix }}%
\providecommand \urlprefix  [0]{URL }%
\providecommand \Eprint [0]{\href }%
\providecommand \doibase [0]{https://doi.org/}%
\providecommand \selectlanguage [0]{\@gobble}%
\providecommand \bibinfo  [0]{\@secondoftwo}%
\providecommand \bibfield  [0]{\@secondoftwo}%
\providecommand \translation [1]{[#1]}%
\providecommand \BibitemOpen [0]{}%
\providecommand \bibitemStop [0]{}%
\providecommand \bibitemNoStop [0]{.\EOS\space}%
\providecommand \EOS [0]{\spacefactor3000\relax}%
\providecommand \BibitemShut  [1]{\csname bibitem#1\endcsname}%
\let\auto@bib@innerbib\@empty
\bibitem [{\citenamefont {Khinchin}(1949)}]{khinchin1949mathematical}%
  \BibitemOpen
  \bibfield  {author} {\bibinfo {author} {\bibfnamefont {A.~I.}\ \bibnamefont {Khinchin}},\ }\href@noop {} {\emph {\bibinfo {title} {Mathematical foundations of statistical mechanics}}}\ (\bibinfo  {publisher} {Courier Corporation},\ \bibinfo {year} {1949})\BibitemShut {NoStop}%
\bibitem [{\citenamefont {Deutsch}(1991)}]{Deutsch1991}%
  \BibitemOpen
  \bibfield  {author} {\bibinfo {author} {\bibfnamefont {J.~M.}\ \bibnamefont {Deutsch}},\ }\bibfield  {title} {\bibinfo {title} {Quantum statistical mechanics in a closed system},\ }\href {https://doi.org/10.1103/PhysRevA.43.2046} {\bibfield  {journal} {\bibinfo  {journal} {Phys. Rev. A}\ }\textbf {\bibinfo {volume} {43}},\ \bibinfo {pages} {2046} (\bibinfo {year} {1991})}\BibitemShut {NoStop}%
\bibitem [{\citenamefont {Srednicki}(1994)}]{Srednicki1994}%
  \BibitemOpen
  \bibfield  {author} {\bibinfo {author} {\bibfnamefont {M.}~\bibnamefont {Srednicki}},\ }\bibfield  {title} {\bibinfo {title} {Chaos and quantum thermalization},\ }\href {https://doi.org/10.1103/PhysRevE.50.888} {\bibfield  {journal} {\bibinfo  {journal} {Phys. Rev. E}\ }\textbf {\bibinfo {volume} {50}},\ \bibinfo {pages} {888} (\bibinfo {year} {1994})}\BibitemShut {NoStop}%
\bibitem [{\citenamefont {Srednicki}(1996)}]{Srednicki1996}%
  \BibitemOpen
  \bibfield  {author} {\bibinfo {author} {\bibfnamefont {M.}~\bibnamefont {Srednicki}},\ }\bibfield  {title} {\bibinfo {title} {Thermal fluctuations in quantized chaotic systems},\ }\href {https://doi.org/10.1088/0305-4470/29/4/003} {\bibfield  {journal} {\bibinfo  {journal} {Journal of Physics A: Mathematical and General}\ }\textbf {\bibinfo {volume} {29}},\ \bibinfo {pages} {L75} (\bibinfo {year} {1996})}\BibitemShut {NoStop}%
\bibitem [{\citenamefont {Page}(1993)}]{Page1993}%
  \BibitemOpen
  \bibfield  {author} {\bibinfo {author} {\bibfnamefont {D.~N.}\ \bibnamefont {Page}},\ }\bibfield  {title} {\bibinfo {title} {Average entropy of a subsystem},\ }\href {https://doi.org/10.1103/PhysRevLett.71.1291} {\bibfield  {journal} {\bibinfo  {journal} {Phys. Rev. Lett.}\ }\textbf {\bibinfo {volume} {71}},\ \bibinfo {pages} {1291} (\bibinfo {year} {1993})}\BibitemShut {NoStop}%
\bibitem [{\citenamefont {\L{}yd\ifmmode~\dot{z}\else \.{z}\fi{}ba}\ \emph {et~al.}(2021{\natexlab{a}})\citenamefont {\L{}yd\ifmmode~\dot{z}\else \.{z}\fi{}ba}, \citenamefont {Zhang}, \citenamefont {Rigol},\ and\ \citenamefont {Vidmar}}]{Vidmar-Rigol2021_PRB-2_1e_ETH}%
  \BibitemOpen
  \bibfield  {author} {\bibinfo {author} {\bibfnamefont {P.}~\bibnamefont {\L{}yd\ifmmode~\dot{z}\else \.{z}\fi{}ba}}, \bibinfo {author} {\bibfnamefont {Y.}~\bibnamefont {Zhang}}, \bibinfo {author} {\bibfnamefont {M.}~\bibnamefont {Rigol}},\ and\ \bibinfo {author} {\bibfnamefont {L.}~\bibnamefont {Vidmar}},\ }\bibfield  {title} {\bibinfo {title} {Single-particle eigenstate thermalization in quantum-chaotic quadratic hamiltonians},\ }\href {https://doi.org/10.1103/PhysRevB.104.214203} {\bibfield  {journal} {\bibinfo  {journal} {Phys. Rev. B}\ }\textbf {\bibinfo {volume} {104}},\ \bibinfo {pages} {214203} (\bibinfo {year} {2021}{\natexlab{a}})}\BibitemShut {NoStop}%
\bibitem [{\citenamefont {\L{}yd\ifmmode~\dot{z}\else \.{z}\fi{}ba}\ \emph {et~al.}(2023)\citenamefont {\L{}yd\ifmmode~\dot{z}\else \.{z}\fi{}ba}, \citenamefont {Mierzejewski}, \citenamefont {Rigol},\ and\ \citenamefont {Vidmar}}]{Vidmar-Rigol2023_PRL_genETH}%
  \BibitemOpen
  \bibfield  {author} {\bibinfo {author} {\bibfnamefont {P.}~\bibnamefont {\L{}yd\ifmmode~\dot{z}\else \.{z}\fi{}ba}}, \bibinfo {author} {\bibfnamefont {M.}~\bibnamefont {Mierzejewski}}, \bibinfo {author} {\bibfnamefont {M.}~\bibnamefont {Rigol}},\ and\ \bibinfo {author} {\bibfnamefont {L.}~\bibnamefont {Vidmar}},\ }\bibfield  {title} {\bibinfo {title} {Generalized thermalization in quantum-chaotic quadratic hamiltonians},\ }\href {https://doi.org/10.1103/PhysRevLett.131.060401} {\bibfield  {journal} {\bibinfo  {journal} {Phys. Rev. Lett.}\ }\textbf {\bibinfo {volume} {131}},\ \bibinfo {pages} {060401} (\bibinfo {year} {2023})}\BibitemShut {NoStop}%
\bibitem [{\citenamefont {Łydżba}\ \emph {et~al.}(2024)\citenamefont {Łydżba}, \citenamefont {Świętek}, \citenamefont {Mierzejewski}, \citenamefont {Rigol},\ and\ \citenamefont {Vidmar}}]{Lydzba2024normal_weak_ETH}%
  \BibitemOpen
  \bibfield  {author} {\bibinfo {author} {\bibfnamefont {P.}~\bibnamefont {Łydżba}}, \bibinfo {author} {\bibfnamefont {R.}~\bibnamefont {Świętek}}, \bibinfo {author} {\bibfnamefont {M.}~\bibnamefont {Mierzejewski}}, \bibinfo {author} {\bibfnamefont {M.}~\bibnamefont {Rigol}},\ and\ \bibinfo {author} {\bibfnamefont {L.}~\bibnamefont {Vidmar}},\ }\href {https://arxiv.org/abs/2404.02199} {\bibinfo {title} {Normal weak eigenstate thermalization}} (\bibinfo {year} {2024}),\ \Eprint {https://arxiv.org/abs/2404.02199} {arXiv:2404.02199 [cond-mat.stat-mech]} \BibitemShut {NoStop}%
\bibitem [{\citenamefont {Vidmar}\ \emph {et~al.}(2017)\citenamefont {Vidmar}, \citenamefont {Hackl}, \citenamefont {Bianchi},\ and\ \citenamefont {Rigol}}]{Vidmar-Rigol2017_PRL-1_TI_Fermi}%
  \BibitemOpen
  \bibfield  {author} {\bibinfo {author} {\bibfnamefont {L.}~\bibnamefont {Vidmar}}, \bibinfo {author} {\bibfnamefont {L.}~\bibnamefont {Hackl}}, \bibinfo {author} {\bibfnamefont {E.}~\bibnamefont {Bianchi}},\ and\ \bibinfo {author} {\bibfnamefont {M.}~\bibnamefont {Rigol}},\ }\bibfield  {title} {\bibinfo {title} {Entanglement entropy of eigenstates of quadratic fermionic hamiltonians},\ }\href {https://doi.org/10.1103/PhysRevLett.119.020601} {\bibfield  {journal} {\bibinfo  {journal} {Phys. Rev. Lett.}\ }\textbf {\bibinfo {volume} {119}},\ \bibinfo {pages} {020601} (\bibinfo {year} {2017})}\BibitemShut {NoStop}%
\bibitem [{\citenamefont {Vidmar}\ and\ \citenamefont {Rigol}(2017)}]{Vidmar-Rigol2017_PRL-2_deviations}%
  \BibitemOpen
  \bibfield  {author} {\bibinfo {author} {\bibfnamefont {L.}~\bibnamefont {Vidmar}}\ and\ \bibinfo {author} {\bibfnamefont {M.}~\bibnamefont {Rigol}},\ }\bibfield  {title} {\bibinfo {title} {Entanglement entropy of eigenstates of quantum chaotic hamiltonians},\ }\href {https://doi.org/10.1103/PhysRevLett.119.220603} {\bibfield  {journal} {\bibinfo  {journal} {Phys. Rev. Lett.}\ }\textbf {\bibinfo {volume} {119}},\ \bibinfo {pages} {220603} (\bibinfo {year} {2017})}\BibitemShut {NoStop}%
\bibitem [{\citenamefont {Hackl}\ \emph {et~al.}(2019)\citenamefont {Hackl}, \citenamefont {Vidmar}, \citenamefont {Rigol},\ and\ \citenamefont {Bianchi}}]{Vidmar-Rigol2019_PRB_XY}%
  \BibitemOpen
  \bibfield  {author} {\bibinfo {author} {\bibfnamefont {L.}~\bibnamefont {Hackl}}, \bibinfo {author} {\bibfnamefont {L.}~\bibnamefont {Vidmar}}, \bibinfo {author} {\bibfnamefont {M.}~\bibnamefont {Rigol}},\ and\ \bibinfo {author} {\bibfnamefont {E.}~\bibnamefont {Bianchi}},\ }\bibfield  {title} {\bibinfo {title} {Average eigenstate entanglement entropy of the xy chain in a transverse field and its universality for translationally invariant quadratic fermionic models},\ }\href {https://doi.org/10.1103/PhysRevB.99.075123} {\bibfield  {journal} {\bibinfo  {journal} {Phys. Rev. B}\ }\textbf {\bibinfo {volume} {99}},\ \bibinfo {pages} {075123} (\bibinfo {year} {2019})}\BibitemShut {NoStop}%
\bibitem [{\citenamefont {\L{}yd\ifmmode~\dot{z}\else \.{z}\fi{}ba}\ \emph {et~al.}(2020)\citenamefont {\L{}yd\ifmmode~\dot{z}\else \.{z}\fi{}ba}, \citenamefont {Rigol},\ and\ \citenamefont {Vidmar}}]{Vidmar-Rigol2020_PRL_SYK2}%
  \BibitemOpen
  \bibfield  {author} {\bibinfo {author} {\bibfnamefont {P.}~\bibnamefont {\L{}yd\ifmmode~\dot{z}\else \.{z}\fi{}ba}}, \bibinfo {author} {\bibfnamefont {M.}~\bibnamefont {Rigol}},\ and\ \bibinfo {author} {\bibfnamefont {L.}~\bibnamefont {Vidmar}},\ }\bibfield  {title} {\bibinfo {title} {Eigenstate entanglement entropy in random quadratic hamiltonians},\ }\href {https://doi.org/10.1103/PhysRevLett.125.180604} {\bibfield  {journal} {\bibinfo  {journal} {Phys. Rev. Lett.}\ }\textbf {\bibinfo {volume} {125}},\ \bibinfo {pages} {180604} (\bibinfo {year} {2020})}\BibitemShut {NoStop}%
\bibitem [{\citenamefont {\L{}yd\ifmmode~\dot{z}\else \.{z}\fi{}ba}\ \emph {et~al.}(2021{\natexlab{b}})\citenamefont {\L{}yd\ifmmode~\dot{z}\else \.{z}\fi{}ba}, \citenamefont {Rigol},\ and\ \citenamefont {Vidmar}}]{Vidmar-Rigol2021_PRB-1_3d_ALT}%
  \BibitemOpen
  \bibfield  {author} {\bibinfo {author} {\bibfnamefont {P.}~\bibnamefont {\L{}yd\ifmmode~\dot{z}\else \.{z}\fi{}ba}}, \bibinfo {author} {\bibfnamefont {M.}~\bibnamefont {Rigol}},\ and\ \bibinfo {author} {\bibfnamefont {L.}~\bibnamefont {Vidmar}},\ }\bibfield  {title} {\bibinfo {title} {Entanglement in many-body eigenstates of quantum-chaotic quadratic hamiltonians},\ }\href {https://doi.org/10.1103/PhysRevB.103.104206} {\bibfield  {journal} {\bibinfo  {journal} {Phys. Rev. B}\ }\textbf {\bibinfo {volume} {103}},\ \bibinfo {pages} {104206} (\bibinfo {year} {2021}{\natexlab{b}})}\BibitemShut {NoStop}%
\bibitem [{\citenamefont {Zanardi}(2002)}]{Zanardi2002}%
  \BibitemOpen
  \bibfield  {author} {\bibinfo {author} {\bibfnamefont {P.}~\bibnamefont {Zanardi}},\ }\bibfield  {title} {\bibinfo {title} {Quantum entanglement in fermionic lattices},\ }\href {https://doi.org/10.1103/PhysRevA.65.042101} {\bibfield  {journal} {\bibinfo  {journal} {Phys. Rev. A}\ }\textbf {\bibinfo {volume} {65}},\ \bibinfo {pages} {042101} (\bibinfo {year} {2002})}\BibitemShut {NoStop}%
\bibitem [{\citenamefont {Peschel}(2003)}]{Peschel2003}%
  \BibitemOpen
  \bibfield  {author} {\bibinfo {author} {\bibfnamefont {I.}~\bibnamefont {Peschel}},\ }\bibfield  {title} {\bibinfo {title} {Calculation of reduced density matrices from correlation functions},\ }\href {https://doi.org/10.1088/0305-4470/36/14/101} {\bibfield  {journal} {\bibinfo  {journal} {Journal of Physics A: Mathematical and General}\ }\textbf {\bibinfo {volume} {36}},\ \bibinfo {pages} {L205} (\bibinfo {year} {2003})}\BibitemShut {NoStop}%
\bibitem [{\citenamefont {Calabrese}\ \emph {et~al.}(2011)\citenamefont {Calabrese}, \citenamefont {Mintchev},\ and\ \citenamefont {Vicari}}]{Calabrese2011}%
  \BibitemOpen
  \bibfield  {author} {\bibinfo {author} {\bibfnamefont {P.}~\bibnamefont {Calabrese}}, \bibinfo {author} {\bibfnamefont {M.}~\bibnamefont {Mintchev}},\ and\ \bibinfo {author} {\bibfnamefont {E.}~\bibnamefont {Vicari}},\ }\bibfield  {title} {\bibinfo {title} {Entanglement entropy of one-dimensional gases},\ }\href {https://doi.org/10.1103/PhysRevLett.107.020601} {\bibfield  {journal} {\bibinfo  {journal} {Phys. Rev. Lett.}\ }\textbf {\bibinfo {volume} {107}},\ \bibinfo {pages} {020601} (\bibinfo {year} {2011})}\BibitemShut {NoStop}%
\bibitem [{\citenamefont {Peschel}\ and\ \citenamefont {Eisler}(2009)}]{Peschel_2009}%
  \BibitemOpen
  \bibfield  {author} {\bibinfo {author} {\bibfnamefont {I.}~\bibnamefont {Peschel}}\ and\ \bibinfo {author} {\bibfnamefont {V.}~\bibnamefont {Eisler}},\ }\bibfield  {title} {\bibinfo {title} {Reduced density matrices and entanglement entropy in free lattice models},\ }\href {https://doi.org/10.1088/1751-8113/42/50/504003} {\bibfield  {journal} {\bibinfo  {journal} {Journal of Physics A: Mathematical and Theoretical}\ }\textbf {\bibinfo {volume} {42}},\ \bibinfo {pages} {504003} (\bibinfo {year} {2009})}\BibitemShut {NoStop}%
\bibitem [{\citenamefont {Dias}\ \emph {et~al.}(2021)\citenamefont {Dias}, \citenamefont {Haque}, \citenamefont {Ribeiro},\ and\ \citenamefont {McClarty}}]{Dias2021}%
  \BibitemOpen
  \bibfield  {author} {\bibinfo {author} {\bibfnamefont {B.~C.}\ \bibnamefont {Dias}}, \bibinfo {author} {\bibfnamefont {M.}~\bibnamefont {Haque}}, \bibinfo {author} {\bibfnamefont {P.}~\bibnamefont {Ribeiro}},\ and\ \bibinfo {author} {\bibfnamefont {P.}~\bibnamefont {McClarty}},\ }\href {https://doi.org/10.48550/ARXIV.2102.09846} {\bibinfo {title} {Diffusive operator spreading for random unitary free fermion circuits}} (\bibinfo {year} {2021})\BibitemShut {NoStop}%
\bibitem [{\citenamefont {Ronzani}\ \emph {et~al.}(2018)\citenamefont {Ronzani}, \citenamefont {Karimi}, \citenamefont {Senior}, \citenamefont {Chang}, \citenamefont {Peltonen}, \citenamefont {Chen},\ and\ \citenamefont {Pekola}}]{Ronzani2018}%
  \BibitemOpen
  \bibfield  {author} {\bibinfo {author} {\bibfnamefont {A.}~\bibnamefont {Ronzani}}, \bibinfo {author} {\bibfnamefont {B.}~\bibnamefont {Karimi}}, \bibinfo {author} {\bibfnamefont {J.}~\bibnamefont {Senior}}, \bibinfo {author} {\bibfnamefont {Y.-C.}\ \bibnamefont {Chang}}, \bibinfo {author} {\bibfnamefont {J.~T.}\ \bibnamefont {Peltonen}}, \bibinfo {author} {\bibfnamefont {C.}~\bibnamefont {Chen}},\ and\ \bibinfo {author} {\bibfnamefont {J.~P.}\ \bibnamefont {Pekola}},\ }\bibfield  {title} {\bibinfo {title} {Tunable photonic heat transport in a quantum heat valve},\ }\href {https://doi.org/10.1038/s41567-018-0199-4} {\bibfield  {journal} {\bibinfo  {journal} {Nature Physics}\ }\textbf {\bibinfo {volume} {14}},\ \bibinfo {pages} {991–995} (\bibinfo {year} {2018})}\BibitemShut {NoStop}%
\bibitem [{\citenamefont {Colpa}(1978)}]{Colpa1978}%
  \BibitemOpen
  \bibfield  {author} {\bibinfo {author} {\bibfnamefont {J.}~\bibnamefont {Colpa}},\ }\bibfield  {title} {\bibinfo {title} {Diagonalization of the quadratic boson hamiltonian},\ }\href {https://doi.org/10.1016/0378-4371(78)90160-7} {\bibfield  {journal} {\bibinfo  {journal} {Physica A: Statistical Mechanics and its Applications}\ }\textbf {\bibinfo {volume} {93}},\ \bibinfo {pages} {327–353} (\bibinfo {year} {1978})}\BibitemShut {NoStop}%
\bibitem [{\citenamefont {Segal}\ and\ \citenamefont {Nitzan}(2005)}]{Segal2005}%
  \BibitemOpen
  \bibfield  {author} {\bibinfo {author} {\bibfnamefont {D.}~\bibnamefont {Segal}}\ and\ \bibinfo {author} {\bibfnamefont {A.}~\bibnamefont {Nitzan}},\ }\bibfield  {title} {\bibinfo {title} {Spin-boson thermal rectifier},\ }\bibfield  {journal} {\bibinfo  {journal} {Physical Review Letters}\ }\textbf {\bibinfo {volume} {94}},\ \href {https://doi.org/10.1103/physrevlett.94.034301} {10.1103/physrevlett.94.034301} (\bibinfo {year} {2005})\BibitemShut {NoStop}%
\bibitem [{\citenamefont {Senior}\ \emph {et~al.}(2020)\citenamefont {Senior}, \citenamefont {Gubaydullin}, \citenamefont {Karimi}, \citenamefont {Peltonen}, \citenamefont {Ankerhold},\ and\ \citenamefont {Pekola}}]{Senior2020}%
  \BibitemOpen
  \bibfield  {author} {\bibinfo {author} {\bibfnamefont {J.}~\bibnamefont {Senior}}, \bibinfo {author} {\bibfnamefont {A.}~\bibnamefont {Gubaydullin}}, \bibinfo {author} {\bibfnamefont {B.}~\bibnamefont {Karimi}}, \bibinfo {author} {\bibfnamefont {J.~T.}\ \bibnamefont {Peltonen}}, \bibinfo {author} {\bibfnamefont {J.}~\bibnamefont {Ankerhold}},\ and\ \bibinfo {author} {\bibfnamefont {J.~P.}\ \bibnamefont {Pekola}},\ }\bibfield  {title} {\bibinfo {title} {Heat rectification via a superconducting artificial atom},\ }\bibfield  {journal} {\bibinfo  {journal} {Communications Physics}\ }\textbf {\bibinfo {volume} {3}},\ \href {https://doi.org/10.1038/s42005-020-0307-5} {10.1038/s42005-020-0307-5} (\bibinfo {year} {2020})\BibitemShut {NoStop}%
\bibitem [{\citenamefont {Ojanen}\ and\ \citenamefont {Jauho}(2008)}]{Teemu2008}%
  \BibitemOpen
  \bibfield  {author} {\bibinfo {author} {\bibfnamefont {T.}~\bibnamefont {Ojanen}}\ and\ \bibinfo {author} {\bibfnamefont {A.-P.}\ \bibnamefont {Jauho}},\ }\bibfield  {title} {\bibinfo {title} {Mesoscopic photon heat transistor},\ }\href {https://doi.org/10.1103/PhysRevLett.100.155902} {\bibfield  {journal} {\bibinfo  {journal} {Phys. Rev. Lett.}\ }\textbf {\bibinfo {volume} {100}},\ \bibinfo {pages} {155902} (\bibinfo {year} {2008})}\BibitemShut {NoStop}%
\bibitem [{\citenamefont {Satrya}\ \emph {et~al.}(2023)\citenamefont {Satrya}, \citenamefont {Guthrie}, \citenamefont {M\"{a}kinen},\ and\ \citenamefont {Pekola}}]{Satrya2023}%
  \BibitemOpen
  \bibfield  {author} {\bibinfo {author} {\bibfnamefont {C.~D.}\ \bibnamefont {Satrya}}, \bibinfo {author} {\bibfnamefont {A.}~\bibnamefont {Guthrie}}, \bibinfo {author} {\bibfnamefont {I.~K.}\ \bibnamefont {M\"{a}kinen}},\ and\ \bibinfo {author} {\bibfnamefont {J.~P.}\ \bibnamefont {Pekola}},\ }\bibfield  {title} {\bibinfo {title} {Electromagnetic simulation and microwave circuit approach of heat transport in superconducting qubits},\ }\href {https://doi.org/10.1088/2399-6528/acbae2} {\bibfield  {journal} {\bibinfo  {journal} {Journal of Physics Communications}\ }\textbf {\bibinfo {volume} {7}},\ \bibinfo {pages} {015005} (\bibinfo {year} {2023})}\BibitemShut {NoStop}%
\bibitem [{\citenamefont {Agarwalla}\ and\ \citenamefont {Segal}(2017)}]{Agarwalla2017}%
  \BibitemOpen
  \bibfield  {author} {\bibinfo {author} {\bibfnamefont {B.~K.}\ \bibnamefont {Agarwalla}}\ and\ \bibinfo {author} {\bibfnamefont {D.}~\bibnamefont {Segal}},\ }\bibfield  {title} {\bibinfo {title} {Energy current and its statistics in the nonequilibrium spin-boson model: Majorana fermion representation},\ }\href {https://doi.org/10.1088/1367-2630/aa6657} {\bibfield  {journal} {\bibinfo  {journal} {New Journal of Physics}\ }\textbf {\bibinfo {volume} {19}},\ \bibinfo {pages} {043030} (\bibinfo {year} {2017})}\BibitemShut {NoStop}%
\bibitem [{\citenamefont {Topp}\ \emph {et~al.}(2015)\citenamefont {Topp}, \citenamefont {Brandes},\ and\ \citenamefont {Schaller}}]{Topp_2015}%
  \BibitemOpen
  \bibfield  {author} {\bibinfo {author} {\bibfnamefont {G.~E.}\ \bibnamefont {Topp}}, \bibinfo {author} {\bibfnamefont {T.}~\bibnamefont {Brandes}},\ and\ \bibinfo {author} {\bibfnamefont {G.}~\bibnamefont {Schaller}},\ }\bibfield  {title} {\bibinfo {title} {Steady-state thermodynamics of non-interacting transport beyond weak coupling},\ }\href {https://doi.org/10.1209/0295-5075/110/67003} {\bibfield  {journal} {\bibinfo  {journal} {Europhysics Letters}\ }\textbf {\bibinfo {volume} {110}},\ \bibinfo {pages} {67003} (\bibinfo {year} {2015})}\BibitemShut {NoStop}%
\end{thebibliography}%
\appendix
\section{Derivation of the anomalous heat current}
\label{app:anomalous_current}
This appendix elaborates the concept of dividing the total heat current into a normal and an anomalous part, which is used in the main text to explain the difference between the exact and the RWA heat currents in the weak-coupling regime. Moreover, the analytical formula for the anomalous heat current given by Eq.~\eqref{eq::anomalous_current} of the main text is derived using perturbation theory.

The Hamiltonian of the single mode heat valve introduced in the main text has the form
\begin{equation}
\begin{split}
    \mathcal{H}/\hbar = &\sum_{\alpha,k}\omega_{\alpha k}a^\dagger_{\alpha k}a_{\alpha k} + \omega_0d^\dagger d+\\&\sum_{\alpha,k} g_{\alpha k}(a_{\alpha k}^{\dagger}d+d^\dagger a_{\alpha k} + a_{\alpha k}^\dagger d^\dagger+da_{\alpha k}),     
\end{split}
\end{equation}
where $\alpha = \{1,2\} $ labels the two baths, $k$ labels the modes in the baths with the frequencies $\{\omega_{\alpha k}\}$, $\omega_0$ is the frequency of the central level connecting the baths and $\{g_{\alpha k}\}$ are the couplings between the central level and each bath mode. The rotating-wave approximation corresponds to dropping the particle non-conserving terms such that
\begin{equation}
\begin{split}
    {\mathcal{H}}_{\mathrm{RWA}}/\hbar
    =&\sum_{\alpha,k}\omega_{\alpha k}a^\dagger_{\alpha k}a_{\alpha k} + \omega_0d^\dagger d+\\&\sum_{\alpha,k} g_{\alpha k}(a_{\alpha k}^{\dagger}d+d^\dagger a_{\alpha k}).    
\end{split}
\end{equation}

The heat current $I^{(\alpha)}(t) = -(i/\hbar)\langle [\mathcal{H_\alpha,\mathcal{H}}]\rangle(t)$ into the bath $\alpha$ reads
\begin{equation}
    I^{(\alpha)}(t) = I_n(t) + I_a(t),  
\end{equation}
where the normal part
\begin{equation}
    I_n^{(\alpha)}(t) = -i\hbar\sum_{k} g_{\alpha k}\omega_{\alpha k} (\langle a_{\alpha k}^\dagger d\rangle - \langle d^\dagger a_{\alpha k}\rangle)
\end{equation}
and the anomalous part
\begin{equation}
    I_a^{(\alpha)}(t) = -i\hbar\sum_{k} g_{\alpha k}\omega_{\alpha k} (\langle a_{\alpha k}^\dagger d^\dagger\rangle - \langle d a_{\alpha k}\rangle).
\end{equation}
In the rotating-wave approximation, the anomalous current is zero at all times regardless of the coupling, whereas in the exact model with sufficiently weak coupling, the anomalous current is responsible for a transient oscillation behavior at short time scales. This is seen by considering the Heisenberg equations of motion. In the case of the exact Hamiltonian, the relevant equations read
\begin{equation}
    \dot{d}(t)=-i\Omega d(t)-i\sum_{\alpha',k'}g_{\alpha'k'}(a_{\alpha'k'}(t)-a^{\dagger}_{\alpha'k'}(t))
\end{equation}
and
\begin{equation}
    \dot{a}_{\alpha k}(t)=-i\omega_{\alpha k}a_{\alpha k}(t)-ig_{\alpha k}(d(t)+d^\dagger(t)),
\end{equation}
which are then conveniently modified into a set of linear integral equations
\begin{equation}
\begin{split}
    d(t)&=d(0)e^{-i\Omega t}\\&-i\sum_{\alpha,k'}g_{\alpha'k'}\int_{0}^{t}dt'(a_{\alpha'k'}(t')e^{i\Omega t'}-a_{\alpha'k'}^\dagger(t')e^{i\Omega t'})e^{-i\Omega t}    
\end{split}
\end{equation}
and
\begin{equation}
\begin{split}
    a_{\alpha k}(t)&=a_{\alpha k}(0)e^{-i\omega_{\alpha k}t}    
    \\&
    -ig_{\alpha k}\int_{0}^{t}dt'(d(t')e^{i\omega_{\alpha k}t'}+d^\dagger(t')e^{i\omega_{\alpha k}t'})e^{-i\omega_{\alpha k}t},
\end{split}
\end{equation}
from which we can immediately calculate the iterative solutions up to first order in the couplings
\begin{equation}
\begin{split}
     d(t)&=d(0)e^{-i\Omega t}
     \\&
    -\sum_{\alpha',k'}g_{\alpha'k'} 
    \left[\frac{e^{-i\omega_{\alpha'k'}t}-e^{-i\omega_0 t}}{\omega_0-\omega_{\alpha' k'}}\right]a_{\alpha' k'}(0)
    \\&
    +\sum_{\alpha',k'} g_{\alpha' k'}\left[\frac{e^{i\omega_{\alpha'k'}t}-e^{-i\omega_0t}}{\omega_0+\omega_{\alpha' k'}}\right]a_{\alpha'k'}^{\dagger}(0)
\end{split}
\end{equation}
and
\begin{equation}
\begin{split}
    a_{\alpha k}(t) &= a_{\alpha k}e^{-i\omega_{\alpha k}t}    
    \\&
    -g_{\alpha k}\left[\frac{e^{-i\omega_0 t}-e^{-i\omega_{\alpha k}t}}{-(\omega_0-\omega_{\alpha k})}\right]d(0)
    \\&
    -g_{\alpha k}\left[\frac{e^{i\omega_0 t}-e^{-i\omega_{\alpha k}t}}{\omega_0+\omega_{\alpha k}}\right]d^{\dagger}(0).
\end{split}
\end{equation}

Therefore, we have
\begin{equation}
    \langle a_{\alpha k}(t)d(t)\rangle
    =
    g_{\alpha k}\frac{1-\langle a_{\alpha k}^{\dagger}(0)a_{\alpha k}(0)\rangle}{\omega_0+\omega_{\alpha k}}(1-e^{-i(\omega_0+\omega_{\alpha k})t})
\end{equation}
and finally combining this with $\langle d^\dagger(t) a_{\alpha k}^\dagger(t)\rangle$ and choosing the initial bath mode occupations according to the Fermi-Dirac distribution $\langle a_{\alpha k}^\dagger(0)a_{\alpha k}(0)\rangle = f(\beta_{\alpha}\hbar\omega_{\alpha k})$ yields an expression for the anomalous current
\begin{equation}
\begin{split}
     I_a^{(\alpha)} &= 2\hbar\sum_{k}g_{\alpha k}^2\omega_{\alpha k}\frac{1-f(\beta_{\alpha}\hbar\omega_{\alpha k})}{\omega_0 + \omega_{\alpha k}}\sin((\omega_0+\omega_{\alpha k})t)   
     \\&
     \approx2\hbar\langle g_{\alpha k}^2\rangle\sum_{k}\omega_{\alpha k}\frac{1-f(\beta_\alpha\hbar\omega_{\alpha k})}{\omega_0+\omega_{\alpha k}}\sin((\omega_0+\omega_{\alpha k})t),
\end{split}
\end{equation}
which is the equation applied in the main text. Furthermore, in the case of the RWA Hamiltonian, the difference is that there is no coupling between the creation and annihilation operators and thus the anomalous heat current must be zero at all times since the chosen initial state contains no anomalous correlations.

\section{Steady-state heat current}
\label{app::steady-state}
This appendix provides the steady-state heat current expressions that are compared with the numerical estimates in the main text. The general steady-state formula valid at arbitrary couplings is due to Ref.~\cite{Topp_2015} and the weak-coupling steady state is according to the standard master equation approach.

In the RWA model, the continuum limit of the steady-state heat current through the single level heat valve is given by the Landauer type formula~\cite{Topp_2015}
\begin{equation}
\label{eq::landauer_formula}
    I_H=\int_{0}^{\infty}\frac{d\omega}{2\pi}|\tau(\omega)|^2\hbar\omega[f(\beta_1\hbar\omega)-f(\beta_2\hbar\omega)],
\end{equation}
where $f(\beta_{\alpha}\hbar\omega)$ is the Fermi-Dirac distribution of the bath $\alpha\in\{1,2\}$ and the transmission coefficient $|\tau(\omega)|^2$ has the form~\cite{Topp_2015}
\begin{equation}
\label{eq::transmission_coefficient}
    |\tau(\omega)|^2 = \frac{\Gamma_1(\omega)\Gamma_2(\omega)}{(\omega-\omega_0-\Sigma(\omega))^2+(\Gamma_1(\omega)/2+\Gamma_2(\omega)/2)^2}
\end{equation}
with the spectral densities of the two baths $\Gamma_{\alpha}(\omega)$ defined by 
\begin{equation}
    \Gamma_{\alpha}(\omega) =  2\pi \sum_{k}|g_{\alpha k}|^2\delta(\omega-\omega_{\alpha k})
\end{equation}
and the total self-energy shift of the central level $\Sigma(\omega) = \Sigma_1(\omega)+\Sigma_2(\omega)$, where the self-energy shift due to interaction with the bath $\alpha \in \{1,2\}$ is calculated via the principal value integral
\begin{equation}
    \Sigma_{\alpha}(\omega) = \frac{1}{2\pi}\mathcal{P}\int_{0}^{\infty}\frac{\Gamma_{\alpha}(\omega')}{\omega-\omega'}d\omega'.
\end{equation}
In the particular case of a uniform bath frequency distribution on the interval $[0,2\omega_0]$, we get 
\begin{equation}
    \Gamma_{\alpha}(\omega) \approx 2\pi\nu_0\langle g_{\alpha k}^2\rangle,
\end{equation}
where $\nu_0 = N/2\omega_0$ is the constant density of states and hence
\begin{equation}
     \Sigma_{\alpha}(\omega) = -\frac{1}{2\pi}\Gamma_\alpha\ln\left(\frac{2\omega_0}{\omega}-1\right),
\end{equation}
which lets us calculate the transmission coefficient $|\tau(\omega)|^2$ according to Eq.~\eqref{eq::transmission_coefficient}, which is finally plugged into Eq.~\eqref{eq::landauer_formula} and the remaining integral is evaluated numerically.

Furthermore, in the limit of weak coupling $\gamma/\omega_0 \ll 1$, the expression of the heat current simplifies to 
\begin{equation}
    I_H = \frac{\Gamma_1\Gamma_2}{\Gamma_1+\Gamma_2}\hbar\omega_0[f(\beta_1\hbar\omega_0)-f(\beta_2\hbar\omega_0)],
\end{equation}

which can also be easily derived from the corresponding Lindblad master equation. In particular, when all the couplings are generated from the uniform interval $[-\gamma/\sqrt{N},\gamma/\sqrt{N}]$, we have $\Gamma_\alpha \propto \langle g_{\alpha k}^2\rangle\propto\gamma^2 $ and therefore $I_H \propto \gamma^2$ in the range of weak coupling $\gamma/\omega_0\ll1$. This dependence changes at stronger coupling strengths, where the finite linewidth of the transmission coefficient $|\tau(\omega)|^2$ can no longer be ignored, which means physically that the central level can also transport heat between the bath levels that are not in resonance with the central level frequency. The eventual failure of the weak-coupling formula is clearly demonstrated in Fig.~\ref{fig::fig2} of the main text.

\end{document}